# Origin of Anomalous Xe-H in Presolar Diamonds: Indications of a "cold" r-process

**Karl-Ludwig Kratz**
*Max-Planck-Institut für Chemie*
*Hahn-Meitner-Weg 1, D-55128 Mainz, Germany*
*and: Department of Physics, University of Notre Dame, USA*
E-mail: k-l.kratz@mpic.de

**Khalil Farouqi**
*Max-Planck-Institut für Chemie*
*Hahn-Meitner-Weg 1, D-55128 Mainz, Germany*
E-mail: k.farouqi@mpic.de

**Ulrich Ott**[1]
*Max-Planck-Institut für Chemie*
*Hahn-Meitner-Weg 1, D-55128 Mainz, Germany;*
*and: University of Western Hungary, Szombathely, Hungary*
E-mail: uli.ott@mpic.de

We report on a concerted effort aimed at understanding the nucleosynthesis origin of Xe-H in presolar nanodiamonds. Previously explored possible explanations have included a secondary neutron-burst process occurring in the He-shell of a type II supernova (SN), as well as a rapid separation, between unstable precursor isobars of a primary r-process, and stable Xe isotopes.
Here we present results from the investigation of a rapid neutron-capture scenario in core-collapse SNe with different non-standard r-process variants. Our calculations are performed in the framework of the high-entropy-wind (HEW) scenario using updated nuclear-physics input. We explore the consequences of varying the wind expansion velocity ($V_{exp}$) for selected electron fractions ($Y_e$) with their correlated entropy ranges (S), and neutron-freezeout temperatures ($T_9$(freeze)) and timescales ($\tau_r$(freeze). We draw several conclusions: For Xe-H a "cold" r-process with a fast freezeout seems to be the favored scenario. Furthermore, eliminating the low-S range (i.e. the "weak" r-process component) and maintaining a pure "main" or even "strong" r-process leads to an optimum overall agreement with the measured $^i$Xe/$^{136}$Xe abundance ratios. Our results can provide valuable additional insight into overall astrophysical conditions of producing the r-process part of the total SS heavy elements in explosive nucleosynthesis scenarios.

# 1. Introduction

Apart from the historical "bulk" solar system (SS) isotopic abundances ($N_\odot$) [1, 2] and the elemental abundances more recently measured for metal-deficient halo stars (for a review, see e.g. [3]), meteoritic grains of stardust, which survived from times before the SS formed (see, e.g. [4-6] represent a third group of "observables" crucial to our understanding of the various nucleosynthesis processes in stars. Within the meteoritic grains, attempts to explain the origin of the nanodiamonds has not progressed as much as the understanding of other types of stardust, such as SiC and oxide grains. A major problem is the small size (average $\simeq$2.6 nm), which does not permit classical single-grain isotopic measurements. Therefore, "bulk" samples (i.e. millions of tiny nanodiamond grains) have to be analyzed. Diagnostic isotopic features are present in several trace elements and suggest a connection to SNe. These include Xe-HL [7] (with enhancements in p- and r-isotopes relative to SS), Kr-H [7] and Pt-H [8, 9] (where the heavy isotopes are enhanced), and Te-H [10] (with a clear enhancement of the r-only isotopes). So far, nucleosynthesis processes suggested to account for the H (heavy, r-process) enhancements include "neutron burst" scenarios of secondary nature [11-13], as well as a regular primary r-process augmented by an "early" separation between final, stable isotopes and radioactive isobars formed during the β-decay back to stability from the initial precursors in the assumed r-process [14]. Since both earlier scenarios are not fully convincing in terms of modern nucleosynthesis conditions, we have initiated a concerted effort to look for isotopic features that may be diagnostic for a standard r-process in high-entropy-wind (HEW) ejecta of core-collapse SNe (for details of our HEW-model, see e.g. [15].

# 2. r-Process Calculations and Xe-H

Production mechanisms for elements beyond Fe by slow (s-process) and rapid (r-process) captures of neutrons, have been known for a long time [2]. However, the search for a robust r-process production site has proven difficult. All proposed scenarios face problems with astrophysical conditions as well as with the necessary nuclear-physics input for isotopes far from β-stability. Among the various models, the SN neutrino-driven or high-entropy wind (HEW) model is one of the best studied mechanisms. However, also for this attractive scenario until recently even in the most sophisticated hydrodynamical models the neutrino-driven wind is proton-rich (electron fraction $Y_e$>0.5) during its entire life, thus precluding r-process nucleosynthesis (see, e.g. [16]).

Instead of assuming uncommon conditions to obtain at least low-entropy neutron-rich ejecta ($Y_e$<0.5) and / or artificially increasing the wind entropy by factors of several (see, e.g. [17]), we have steadily optimized our SN HEW parameter studies with simplified dynamics using the computational method(s) and basic nuclear-physics input as outlined in [15]. In fact, this seems to be justified by new investigations on charged-current neutrino interaction rates, which may significantly alter the earlier proton-rich $Y_e$ towards more neutron-rich conditions (see, e.g. [18]).

Commonly, r-process nucleosynthesis calculations are performed to reproduce the SS r-process pattern (see, e.g., [1, 15, 19]), historically defined as the abundance differences between the total SS abundances [2] and the assumed s-process contribution for SS metallicity [20] -- $N_{r,\odot} \simeq N_\odot - N_{s,\odot}$. Such comparisons of isotopic abundance patterns require best possible accuracy of both astrophysical and nuclear-physics parameter input. Given the progress during the past three decades on modeling and measuring nuclear properties near and in the r-process boulevard (see, e.g, [19, 21-24]), in particular for nuclei involved in the creation of the $A \simeq 130$ $N_{r,\odot}$ peak, we have performed an exploratory study of possible modifications of the "standard" r-process (for SS abundances) that can reproduce the observed peculiar isotopic pattern of Xe-H in presolar nanodiamonds [7, 9, 14]. In these stardust samples, the Xe-H abundance pattern is given as ratios relative to the assumed "r-only" isotope $^{136}$Xe, and have (somewhat model-dependent) values of $^{129,131,132,134}$Xe/$^{136}$Xe = 0.207, 0.178, 0.167, 0.699 [9, 14], which are clearly different from the $N_{r,\odot}$ abundance ratios of 3.37, 2.55, 2.24, 1.18 [20].

As for the nuclear-physics input used in our calculations, all theoretical properties are consistently based on the deformed, quenched mass model ETFSI-Q [22]. Using its $Q_\beta$ and $S_n$ values as well as the predicted ground-state shape, β-decay half-lives ($T_{1/2}$) and β-delayed neutron branching ratios ($P_{xn}$) were calculated within the QRPA approach as outlined in [21]. Compared to [15], two additional local improvements in the $A \simeq 130$ peak region [24] were added. First is the inclusion of the $T_{1/2}$ of known and theoretical $\pi p_{1/2}$ isomers in addition to the $\pi g_{9/2}$ ground-state decays in the N = 82 odd-proton nuclei from $^{131}$In to $^{123}$Tc. The second corresponds to new QRPA(GT+ff) calculations for the mass region of exotic $83 < N < 87$ r-process isotopes between Mo and Cd, which are based on optimized model parameters to reproduce experimental level structures in $^{130}$In and $^{131}$In. The improvement in reproducing the "bulk" $N_{r,\odot}$ isotopic abundances in the $A \simeq 130$ peak obtained by these updates is shown, for example, in Fig. 5 of [24]. It is worth emphasizing at this point that a satisfactory agreement between observations and calculations for bulk SS abundances is essential for the interpretation of measured abundance patterns deviating from the $N_{r,\odot}$ distribution.

All results shown in the following were obtained using our parameterized SN HEW model, as outlined in [15]. We have started our study using the correlated "standard" astrophysics parameters (according to our "r-process strength function", where the ratio of free neutrons to seed nuclei is given as $Y_n/Y_{seed} \simeq k_{SN} \times V_{exp} \times (S/Y_e)^3$) with an electron fraction of $Y_e$=0.45, an expansion velocity of the ejecta of $V_{exp}$ = 7500 km/s and an entropy range of 20<S<280 $k_B$/baryon. Under these conditions, we have a "hybrid" r-process type with a freezeout temperature for the free neutrons of $T_9$(freeze) $\simeq$ 0.82, which is reached at a time $\tau_r$(freeze) $\simeq$ 138 ms after the r-process seed formation. This r-process variant reaches the maximum abundance at the top of the $A \simeq 130$ r-abundance peak at S≃195 for $Y_n/Y_{seed} \simeq 35$ with the peak abundance at A = 128. The corresponding $^{129}$Xe/$^{136}$Xe abundance ratio, taken here as representative, is 2.77, which is close to the $N_{r,\odot}$ ratio of 3.37. This result is not sensitive to the assumed $Y_e$ value (i.e. the neutron-richness of the wind ejecta). We found that, compared to the above abundance ratio for $Y_e=0.45$, this value only increases by about 12 % by going down to $Y_e=0.40$ or up to $Y_e=0.48$, respectively.

Given this, for the further calculations we have kept the electron fraction constant at $Y_e=0.45$ and studied the influence of the expansion velocity in the range $1000<V_{exp}<30,000$ km/s, together with the other parameters correlated according to the above defined "strength function", while still keeping the full entropy ranges. For a low expansion velocity of the ejecta of $V_{exp} = 1000$ km/s we obtain a so-called "hot" r-process variant with a relatively long neutron freezeout time of $\tau_r(freeze) \simeq 560$ ms and a freezeout temperature of $T_9(freeze) \simeq 1.3$. The top of the second r-process peak lies at A = 130, for a high $S \simeq 370$ $k_B$/baryon. In contrast, for a high $V_{exp}$ of 30,000 km/s we obtain a fast, "cold" r-process variant with $\tau_r(freeze) \simeq 55$ ms, $T_9(freeze) \simeq 0.4$. The top of the second r-process peak is at A = 127 for a low S of $\simeq 120$ $k_B$/baryon. As can be seen from Fig. 1, which shows the results for a number of selected expansion velocities of the wind ejecta ($2000<V_{exp}<18,000$ km/s), none of these parameter combinations with the respective full entropy ranges is able to reproduce in a satisfactory manner all $^i$Xe/$^{136}$Xe ratios measured in the presolar nanodiamond samples. However, it is apparent that the "best" overall agreement with the observations will very likely be obtained under "cold" r-process conditions. Therefore, in the last step of our parameter study we focused on the parameter combinations of such "non-standard" $10,000<V_{exp}<30,000$ km/s r-process variants, and studied the effect of choosing different entropy ranges.

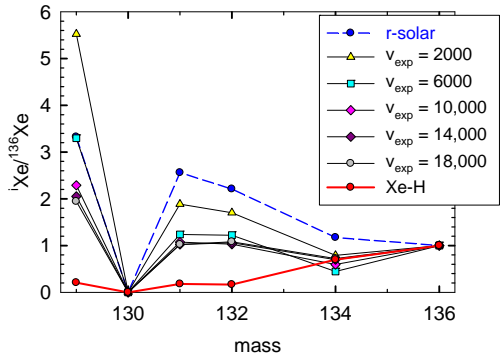 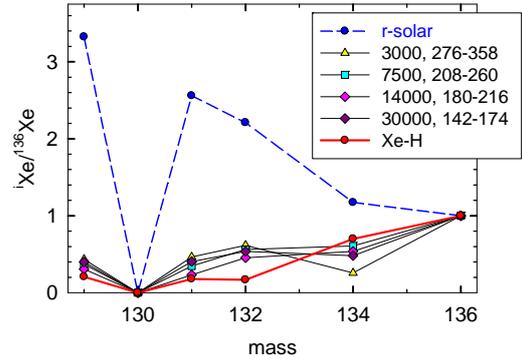

Figure 1: HEW predictions of $^i$Xe/$^{136}$Xe abundance ratios for neutron-rich r-process ejecta with an electron fraction $Y_e = 0.45$ as a function of expansion velocity $V_{exp}$ (as given in the legend in km/s units) for the full entropy ranges. The $N_{r\odot}$ residuals and Xe-H data are included for comparison.

Figure 2: HEW predictions of $^i$Xe/$^{136}$Xe-abundance ratios for constant $Y_e = 0.45$ and different expansion velocities $V_{exp}$ [km/s] with the respective "best-fit" entropy ranges Delta-S [kB/baryon], as given in the legend. For comparison, again $N_{r\odot}$ residuals and Xe-H data included.

As we pointed out earlier already [9, 25], for given $Y_e$ and $V_{exp}$, the different Xe isotopes are produced in different entropy ranges. For our standard conditions used here, the "lightest" Xe-H nuclide $^{129}$Xe is formed most effectively at $S \simeq 185$ $k_B$/baryon, at a slightly lower entropy than required to form the $N_{r,\odot}$ A=130 top of the peak. The heavier isotopes follow at higher S values "beyond" the peak-top conditions. Finally, $^{136}$Xe has its most effective production at $S \simeq 220$ $k_B$/baryon, where already the REE pygmy-peak region and part of the third $N_{r,\odot}$ peak are formed (see, e.g. Fig. 8 in [15]). This suggests that the "low" $^i$Xe/$^{136}$Xe ratios observed in the nanodiamonds can be reproduced when limiting the HEW ejecta to a "main" or even "strong" r-process component with "high" entropies; i.e. in principle by cutting out from the entropy parameter space the lower-S "weak" r-process component.

Thus, for our "standard hybrid" $Y_e$=0.45, $V_{exp}$ = 7500 km/s conditions, a cut of the low-S "weak" r-process component (S<185 $k_B$/baryon) leads to a considerable reduction of the $^{129}$Xe abundance but has only a small effect on $^{136}$Xe. This results in a decrease of the original, full-S range abundance ratio of $^{129}$Xe/$^{136}$Xe = 2.77 by a factor 3 to 0.923. Under these HEW conditions, the "best" possible fit to the Xe-H pattern is obtained for a "strong" r-process variant with an entropy cut at S≃208 $k_B$/baryon, with abundance ratios $^{129,131,132,134}$Xe/$^{136}$Xe = 0.370, 0.347, 0.561, 0.608, respectively. Although $^{129}$Xe, $^{131}$Xe and $^{132}$Xe are still too high by factors 1.8, 1.9 and 3.4, respectively, $^{134}$Xe is met within 15 % to the observed ratio.

As expected, even better agreement with the Xe-H pattern is obtained for "cold" r-process variants in the range 10,000<$V_{exp}$<20,000 km/s, with corresponding entropy cuts of S>206 to S>162 $k_B$/baryon, respectively. A selection of "best fit" examples is shown in Fig. 2, which shows that a "strong cold" r-process can indeed reproduce the Xe-H abundance ratios of $^{129,131,134}$Xe/$^{136}$Xe within 20 to 30 %. Only $^{132}$Xe/$^{136}$Xe, with a "best" ratio of 0.436, remains too high (by a factor ~2.6). In this special case, the remaining deviation may be due to a non-optimum nuclear-physics input. Whereas for the lighter stable Xe isotopes most of the nuclear parameters are either experimentally available or are reasonably well understood by short-range extrapolations, the region beyond A ≃ 133 is rather uncertain, not so much in terms of $T_{1/2}$ values, but rather in the individual branching ratios of β-delayed neutrons. Their measurements clearly will be a challenge for future experiments.

Interestingly, preliminary results for platinum indicate that the HEW conditions found to be favorable for Xe-H can also account for the tentatively reported Pt-H in presolar nanodiamonds. With Xe-H lying in the A ≃ 130 $N_{r,\odot}$ peak and Pt-H in the A ≃ 195 peak, these cosmochemical samples with isotopic abundance patterns can provide constraints on the astrophysical conditions for the production of a full r-process, which cannot be deduced with this sensitivity by the elemental abundance patterns of metal-poor halo stars.